\documentstyle[12pt,epsfig]{article}

\def\shiftdown#1{#1\llap{\lower.04ex\hbox{#1}}}

\begin{document}




\vspace*{0.3cm}

\begin{center}
{\bf {\large 
Relativistic field-theoretical formulation
of the three-dimensional equations
for the three fermion system.
\footnotemark}}

\vspace{0.7cm}
\end{center}

\footnotetext{
Presented at NAPP2003 Conference, Dubrovnik, Croatia,
25.05-01.06.2003.}

\vspace{0.7cm}

\begin{center}
{\large {\it A.\ I.\ Machavariani$^{\diamond \ast}$ }}

{\em $^{\diamond }$ Joint\ Institute\ for\ Nuclear\ Research,\ Moscow
Region \\
141980 Dubna,\ Russia}\\[0pt]
{\em $^{*}$ High Energy Physics Institute of Tbilisi State University,
University Str. 9 }\\[0pt]
{\em 380086 Tbilisi, Georgia }
\end{center}

\vspace{0.5cm} \medskip

\begin{abstract}
A new kind of the relativistic three-body 
 equations for the three fermion systems
 are suggested.  
These equations are derived in
the framework of the standard field-theoretical $S$-matrix approach in the
time-ordered three dimensional form. Therefore corresponding relativistic
covariant equations are three-dimensional from the beginning and  the
considered formulation is free of the ambiguities which appear due to a
three dimensional reduction of the four dimensional Bethe-Salpeter
equations. The solutions of the considered equations satisfy 
automatically the unitarity
condition and for the leptons these equations
are exactly gauge invariant even after the truncation over the
multiparticle ($n>3$) intermediate states. Moreover, the form of these
three-body equations does not depend on the choice of the model Lagrangian
and it is the same for the formulations with and without quark degrees of
freedom. The effective potential of the suggested equations is defined by
the vertex functions with two on-mass shell particles. It is emphasized that
these INPUT vertex functions can be constructed from experimental data.

Special attention is given to the comparison with the three-body
Faddeev equations. Unlike to these equations, the suggested three-body
equation have the form of the Lippmann-Schwinger-type equations with the 
connected potential. In addition, the microscopical potential of the
suggested equations contains 
the contributions from the three-body forces and 
from the particle creation (annihilation) mechanism on the
one external particles. The structure of the three-body 
forces, appearing in the considered field-theoretical formulation, is analyzed.

\end{abstract}

\newpage

\begin{center}
{\bf 1. INTRODUCTION}
\end{center}

\medskip

The purpose of my talk is to focus on the three-body
equations with  the particle creation and annihilation  
phenomena. The most popular tool for this investigation is the three-body 
generalization of the Bethe-Salpeter equations \cite{Kvin,Afnan}.
Unfortunately in this four-dimensional formulation arise a set of 
complications which 
result in a serious approximations.
 For instance, by the three-dimensional 
reduction of the four-dimensional Bethe-Salpeter equations arise the 
ambiguities with the choice of the form of the three-dimensional Green 
functions and the three-dimensional effective potentials.
 Next, the difficults with the unitarity and the gauge invariance 
force to use the tree approximation for the effective 
potentials and the one-particle propagators
by  the practical calculation. In
addition the potential of the Bethe-Salpeter equation
is constructed through the  three-variable vertex functions,
which  are
required as the ``input''  functions. Therefore, in the
practical calculations based on the Bethe-Salpeter equations or their
quasipotential reductions the off-mass shell variables in the vertex
functions are usually neglected or a separable form for all three variables
is introduced.

In my talk I shall consider the other way of derivation of the 
three-body field-theoretical equations
 which allows us to avoid  the above difficults and which 
can be solved with the considerable less number of approximations.  
The organization of this talk is as follows. 
As first I will consider the
three-body spectral decomposition equations (which have the form of the off
shell unitarity conditions \cite{New,Gold}) for the amplitudes of the 
three fermion systems. These equation form a base for the derivation of the 
Lipmann-Schwinger type equation \cite{Gold,New}.
After separation of the
 connected and disconnected parts in the amplitudes and effective 
potentials in these 
three-body spectral decomposition equations, one can separate the 
three-body equations
for the connected and
disconnected parts in the three-body amplitudes.
Next after linearization of these three-body equations 
 we will get the three-body Lipmann-Schwinger equations
for the connected part of the three-body amplitudes.
The major  difference between these equations and the Faddeev equations 
will be discussed. 
Afterwards,  I shall consider 
the structure of the three-body 
potentials for the three-fermion (three  electron or three nucleon,
or for the ed scattering etc.)
systems. Finally the short summary will be
presented.

\begin{center}
{\bf 2.  The three-body Lippmann-Schwinger type equations
for the three-fermion scattering reactions
}
\end{center}

\medskip

The problem of the relativistic description of an particle interactions in
the framework of a potential picture is usually solved by relativistic
generalization of the Lippmann-Schwinger type equation of the
nonrelativistic collision theory \cite{New,Gold}. 
As for basis  for the derivation of the 
Lippmann-Schwinger type equations in the collision theory,
one can  use 
the following quadratically nonlinear
integral equations \cite{Gold}

$$
T_{\alpha\beta}(E_{\beta})=V_{\alpha\beta}+ 
\sum_{\gamma}T_{\alpha\gamma}(E_{\gamma})
{1\over{{E_{\beta}-E_{\gamma}+i\epsilon }}} T_{\beta \gamma}^{\ast}
(E_{\gamma})
+\sum_{d3}T_{\alpha,d3}(E_{d3})
{1\over{{E_{\beta}-E_{d3} }}} T_{\beta,d3}^{\ast}(E_{d3}),
\eqno(2.1)$$
where 
$T_{\alpha\beta}(E_{\beta})$ is the transition amplitude
between the channels $\alpha$ and $\beta$. The 
 $\alpha,\beta,\gamma$ denotes the pure three noninteracting
fermion channels or 
or the one-fermion$+$two-body cluster states. For example for the three
nucleon systems $\alpha=NNN\ or Nd$, for the two electron and nucleon
$\alpha=eeN\ or\ ed$ etc., $d3$ relates to the three-body bound states. 
 $\sum_{\gamma}$
stands for the integration over the momenta and the summation over the
quantum numbers of the complete set intermediate 
$\gamma\equiv |\gamma>$-channel states.

If we suppose, that there exists the full hermitian Hamiltonian
$H$ which has the complete set of 
the eigenfunctions 
$H|\Psi_{\gamma}>= E_{\gamma}|\Psi_{\gamma}>$,
then one can easy reduce eq. (2.1) to the 
Lippmann-Schwinger type equations

$$
T_{\alpha\beta}(E_{\beta})
=V_{\alpha\beta}+\sum_{\gamma}V_{\alpha\gamma}{\frac{1}{{E_{\beta}-
E_{\gamma}+i\epsilon}}}T_{\gamma\beta}(E_{\beta}),\eqno(2.2)
$$

where we have used  the decomposition formula of the full
Green function $G(E)=1/(E-H+i\epsilon)$ over the complete set of the
functions $|\Psi_{\gamma}>$ and we have taken into account the
 connection formula  between a amplitude, a multichannel potential
and a wave function
$T_{\alpha\beta}(E_{\beta})=
<\alpha|V|\Psi_{\beta}>$ \cite{New,Gold}.

The three-body equations (2.1)  are well defined after separation
of the connected ($T_{\alpha\beta}^c$;  $V _{\alpha\beta}^c$)
and disconnected  ($V_{\alpha\beta}^{dc}$; $V _{\alpha\beta}^{dc}$)
parts of amplitudes and potentials. Therefore we split the complete amplitude
and the complete potential in Eq.(2.1) into two corresponding parts

$$V_{\alpha\beta}=V_{\alpha\beta}^{c}+V_{\alpha\beta}^{dc};\ \ \ 
T_{\alpha\beta}(E_{\beta})=
T_{\alpha\beta}^{c}(E_{\beta})+T_{\alpha\beta}^{dc}(E_{\beta}).\eqno(2.3)
$$

 The disconnected parts of the two-body and three-body amplitudes
are depicted in Fig. 1. The disconnected part of the three-body amplitudes 
are independent from the connected part of these amplitudes, because
the two-body clusters with the asymptotic free third particle
is independent on the three-particle interacted clusters. This is 
the requirement
of the independency of the 
asymptotic clusters. The independence of the equations 
for the disconnected part of Eq. (2.1) can be easy demonstrated
in the quantum-field formulation \cite{M7}. As result equation (1) is splited
into two set of independent equations  

$$
T_{\alpha\beta}^{dc}(E_{\beta})=V_{\alpha\beta}^{dc}+ 
\sum_{\gamma}T_{\alpha\gamma}^{dc}(E_{\gamma})
{1\over{{E_{\beta}-E_{\gamma}+i\epsilon }}} \Bigl[T_{\beta \gamma}^{dc}
(E_{\gamma})\Bigr]^{\ast},
\eqno(2.4)$$

$$
T_{\alpha\beta}^c(E_{\beta})=W_{\alpha\beta}+ 
\sum_{\gamma}T_{\alpha\gamma}^c(E_{\gamma})
{1\over{{E_{\beta}-E_{\gamma}+i\epsilon }}} \Bigl[T_{\beta \gamma}^c
(E_{\gamma})\Bigl]^{\ast}
+\sum_{d3}T_{\alpha,d3}(E_{d3})
{1\over{{E_{\beta}-E_{d3} }}} T_{\beta,d3}^{\ast}(E_{d3}),
\eqno(2.5)$$

where

$$W_{\alpha\beta}=V_{\alpha\beta}^c+
\sum_{\gamma}T_{\alpha\gamma}^c(E_{\gamma})
{1\over{{E_{\beta}-E_{\gamma}+i\epsilon }}} \Bigl[T_{\beta \gamma}^{dc}
(E_{\gamma})\Bigl]^{\ast}
+\sum_{\gamma}T_{\alpha\gamma}^{dc}(E_{\gamma})
{1\over{{E_{\beta}-E_{\gamma}+i\epsilon }}} \Bigl[T_{\beta \gamma}^c
(E_{\gamma})\Bigl]^{\ast}.\eqno(2.6)$$



\begin{figure}[htb]
\centerline{\epsfysize=145mm\epsfbox{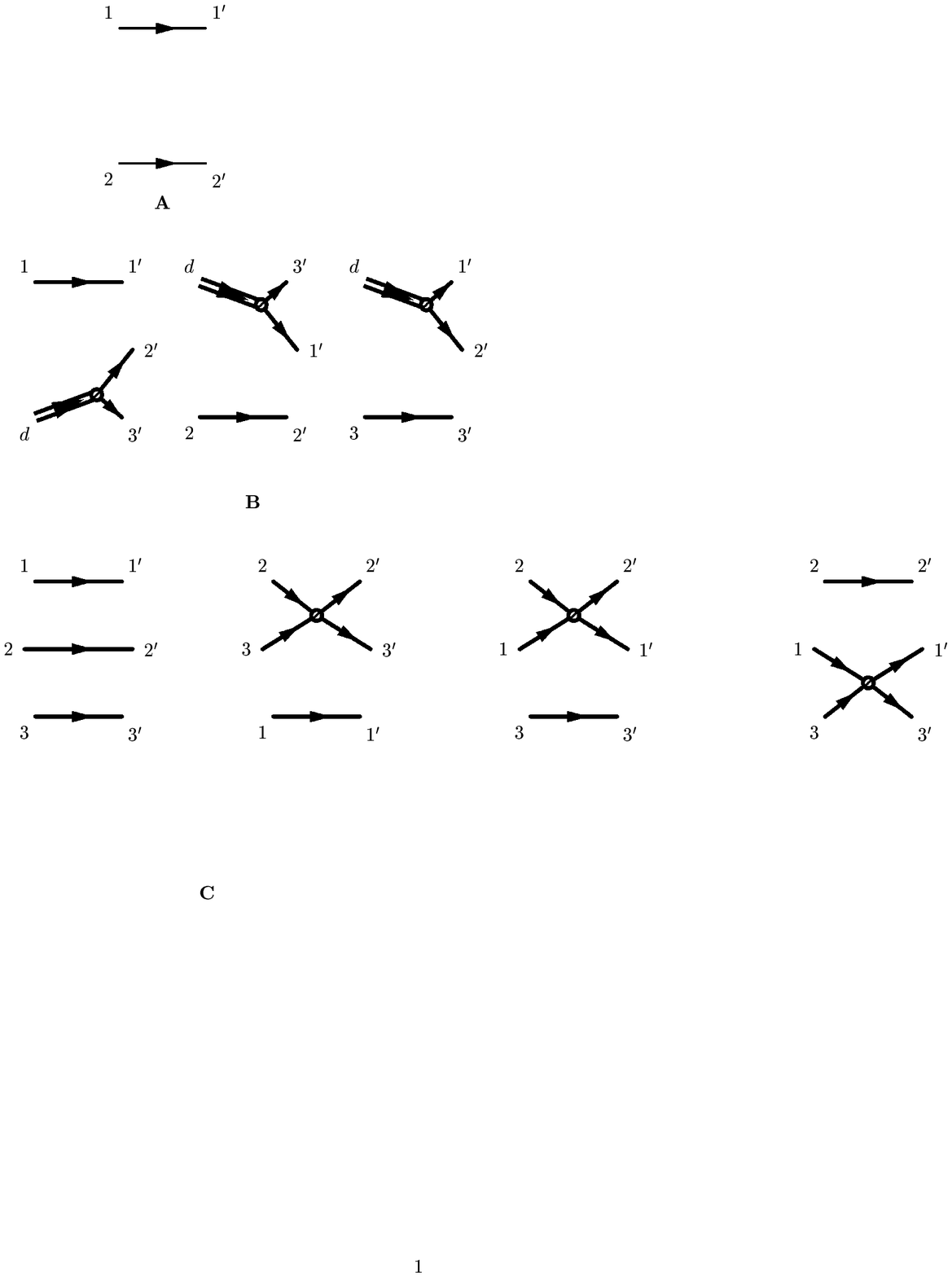}} 
\vspace{-4.0cm} 
\caption{{\protect\footnotesize {\it The disconnected parts of the 
two-fermion 
 and the three-fermion $S$-matrix elements
correspondingly in Fig.1A and Fig.1B, Fig.1C. 
The label $d$ stands for the two-fermion bound state.
The shaded circle corresponds to the two-body scattering amplitude.
}}}
\label{fig:one}
\end{figure}
\vspace{5mm}

The effective potential (2.6) in the field-theoretical
formulation $w_{\alpha,\beta}$  
is not hermitian due to the particle propagators in the 
intermediate states. 
Nevertheless, we have shown in ref.\cite{M1,M7} 
that quadratically nonlinear equations (2.5) are
equivalent to the following Lippmann-Schwinger type equations

$$
{\cal T}_{\alpha\beta}(E_{\beta})=U_{\alpha\beta}(E_{\beta})+
\sum_{\gamma} U_{\alpha\gamma}(E_{\beta}) { 1\over
{ E_{{ \beta} }-E_{\gamma}+i\epsilon } } {\cal T}_{
\gamma\beta}(E_{\beta}),\eqno(2.7)
$$

where
for the sake of simplicity we have omitted the delta function 
 for the total
three-momentum conservation 
$\delta$ function $(2\pi)^3\delta({\bf %
P_{\beta}-P_{\gamma} })$ in above equations.

The explicit form of the linear energy depending potential

$$
U_{\alpha \beta }(E)=A_{\alpha \beta }+E\ B_{\alpha \beta }\eqno(2.8a)
$$
with hermitian $A$ and $B$ matrices 
$$
A_{\alpha \beta }=A_{\beta \alpha }^{*};\ \ \ \ \ \ \ B_{\alpha \beta
}=B_{\beta \alpha }^{*},\eqno(2.8b)
$$
is considered in the next section. 
$U_{\alpha \beta }(E)$ is simply connected
with the $W_{\alpha \beta }$-potential (2.6)

$$
U_{\alpha \beta }(E_{\alpha })=W_{\alpha \beta }.\eqno(2.9a)
$$
Therefore, for any potential $W_{\alpha \beta }$ one
can unambiguously construct $U_{\alpha \beta }(E)$.

Solutions of the equations (2.5) and (2.7) 
coincide on energy shell 
$$
{\cal T}_{\alpha\beta}(E_{\beta}=E_{\alpha})=T_{\alpha\beta}^c
|_{_{_{E_{\beta}=E_{\alpha} } } }\eqno(2.9b)
$$
and in the half on energy shell region these amplitudes are 
simply connected

$$
{\cal T}_{\alpha\beta}=W_{\alpha\beta}+\sum_{\gamma}
W_{\alpha\gamma} {\frac{1}{{E_{\beta}-E_{\gamma}+i\epsilon}}} 
T_{\gamma\beta}^c(E_{\beta}) .\eqno(2.10)
$$

The Lippmann-Schwinger type equations (2.7) 
are our final equations for the three-fermion
scattering amplitudes.
From the other hand, Eq. (2.1) can be  linearized without 
the separation of 
the connected and disconnected parts and from the 
Lippmann-Schwinger equation (2.2) one can derive the Faddeev type equations
\cite{New}. The advantage of Eq. (2.7) is that it does 
 not need the 
splitting into four pieces $V_{\alpha\beta}=\sum_{i=1}^3V_{\alpha\beta}^i+
V_{\alpha\beta}^c$
in order to taken into account the disconnected 
parts in the perturbation series.  Besides Eq. (2.7) are free from the 
overcaunting problem which appear due to special disconnected diagrams
and which generates the corresponding modification of the effective 
potentials\cite{Kvin,Afnan}.

\medskip

\begin{center}
{\bf 3.  
 The three-dimensional three-body field-theoretical equations}
\end{center}

\medskip

In the standard formulation of the quantum field theory \cite{BD,IZ,GrossB}
the $S$-matrix element between the asymptotic three-body 
states $\alpha=1',2',3',f'd'$ and 
$\beta=1,2,3,fd$ 
is connected with the scattering
amplitude $f_{\alpha,\beta}$ 

$$
S_{\alpha,\beta}=<out;\alpha|\beta;in>
=<out;{\widetilde \alpha}|b_{\bf p_a}(in)|{\widetilde \beta};in>
-(2\pi)^4i\delta^{(4)}(P_\alpha-P_\beta) 
f_{\alpha,\beta}\eqno(3.1)
$$

where $f$ denotes the one-fermion state, $d$ stands for the bound
state of two fermions,
$P_{\alpha}\equiv (E_{\alpha},{\bf P}_{\alpha})$ is the
complete four-momentum of the asymptotic state $\alpha$,
$a$ and $b$ corresponds to the one-fermion states extracted
from the asymptotic $\alpha$ and $\beta$ states
 
$$\alpha=a+{\widetilde \alpha};\ \ \ 
\beta=b+{\widetilde \beta}\eqno(3.2)$$

and the four-momentum of the asymptotic one-fermion states $a$ is 
$p_{a}=\Bigl(\sqrt{m_{a}^2+{\bf p}_{a}^2},{\bf p}_{a}\Bigr) \equiv
\Bigl( E_{\bf p_a},{\bf p}_{a}\Bigr)$. 
The amplitude $f_{\alpha\beta}$ has the form 

$$f_{\alpha\beta}=-<out;{\widetilde \alpha}|J_{\bf p_a}(0)|\beta; in>
\eqno(3.4)$$

where $J_{\bf p_a}(x)$ is the current operator of the fermion $a$
which is determined by the Dirac equation
$J_{\bf p_a}(x)=Z_a^{-1/2}{\overline u}({\bf p_a})(i\gamma_{\mu}
\partial_x^{\mu}-m_a)\psi_a(x)$ 
with the renormalization constant $Z_a$ and Dirac bispinor 
function $u({\bf p_a})$ \cite{BD,IZ}.

Using the well know  reduction formulas 
we obtain

$$f_{\alpha\beta}=
<out;{\widetilde \alpha}|b_{\bf p_b}^+(out)J_{\bf p_a}(0)
|{\widetilde \beta};in>$$ 
$$
-<out;{\widetilde \alpha}|\Bigl\{J_{\bf p_a}(0),b_{\bf p_b}^+(0)\Bigr\} 
|{\widetilde \beta};in> 
+i\int d^4 x e^{-ip_b x}<out;
{\widetilde\alpha} |T\Bigl(
J_{\bf p_a}(0){\overline J}_{\bf p_b}
(x)\Bigr) |{\widetilde \beta};in>,\eqno(3.5)
$$

where

$$
b_{{\bf p}_{b}}^+(x_0)=Z_b^{-1/2} \int d^3x e^{-ip_{b}x} 
{\overline u}({\bf p_b})
\gamma_o {\psi}_{b}(x),\eqno(3.6)
$$
Here and afterwards
we use the definitions and normalization conditions from the Itzykson and
Zuber's book \cite{IZ}.

After substitution of the complete
set of the asymptotic states $''in''$ states 
$\sum_n
|n;in><in;n|={\widehat 1}$ between the current operators in expression 
(3.5) and after integration over $x$ we get

$$
f_{\alpha\beta}= W_{\alpha\beta}+(2\pi)^3\sum_{\gamma} f_{\alpha\gamma} {%
\frac{{\delta^{(3)}( {\bf p}_b+{\bf P_{{\widetilde \beta}}-P_{\gamma} })}}{{%
E_{\bf p_b}+{\ P_{{\widetilde \beta}}^o-P_{\gamma}^o+i\epsilon }} }} 
{{\cal T}^{\ast}}_{\beta \gamma} \eqno(3.7)
$$

where
$W_{\alpha\beta}$ contains all contributions of the intermediate
states 
 that can be appear
in the $\beta\to \alpha$ reaction except the 
$s$-channel three particle $\gamma=1''2''3''$ exchange states
and one-fermion $f$
$+$ two-fermion bound states $d$ 
($\gamma=f+d$)
 exchange terms 
which are included  in the second part of  Eq.(3.7).

$$
W_{\alpha\beta}= 
-<out;{\widetilde \alpha}|\Bigl\{J_{\bf p_a}(0),b_{\bf p_b}^+(0)\Bigr\} 
|{\widetilde \beta};in> $$
$$+(2\pi)^3\sum_{n=1''2''3''b'',f''d'',...}<out;{\widetilde\alpha} |
J_{\bf p_a}(0)|n;in>
{ {\delta^{(3)}( {\bf p}_b+{\bf P}_{\widetilde \beta}-{\bf P}_{n} ) }\over
{E_{\bf p_b}+P_{\widetilde \beta}^o-P_{n}^o+i\epsilon } }
<in;n|{\overline J}_{\bf p_b}(0)|{\widetilde \beta};in>
$$
$$
-(2\pi)^3\sum_{l=f,fb,...} 
<out;{\widetilde \alpha}| {\overline J}_{\bf p_b}(0)|l;in>
 { {\delta^{(3)}( {-{\bf p}}_b+{\bf P_{\widetilde \alpha}-P_{l} }) } 
\over{-E_{\bf p_b}+ P_{{\widetilde \alpha}}^o-P_{l}^o } }
<in;l|J_{\bf p_a}(0)|{\widetilde \beta};in>,\eqno(3.8)
$$
where $n=1''2''3''b'',f''d'',...$ denotes the four body 
 states with the intermediate boson $b''$ which denotes
a photon for the three leptons system and $b''$ stands for
 the intermediate
meson for the three barion systems. 
The third part of Eq.(3.8) describes the $u$-channel 
interaction terms which are obtained after crossing of the $a$ and
$b$ particles. The intermediate states of this term contain 
one fermion $l=f$ and one-fermion$+$boson $l=f+b$ states.
These diagrams are depicted in Fig. 2A and in Fig. 2E.


\vspace{5mm}


\begin{figure}[htb]
\centerline{\epsfysize=145mm\epsfbox{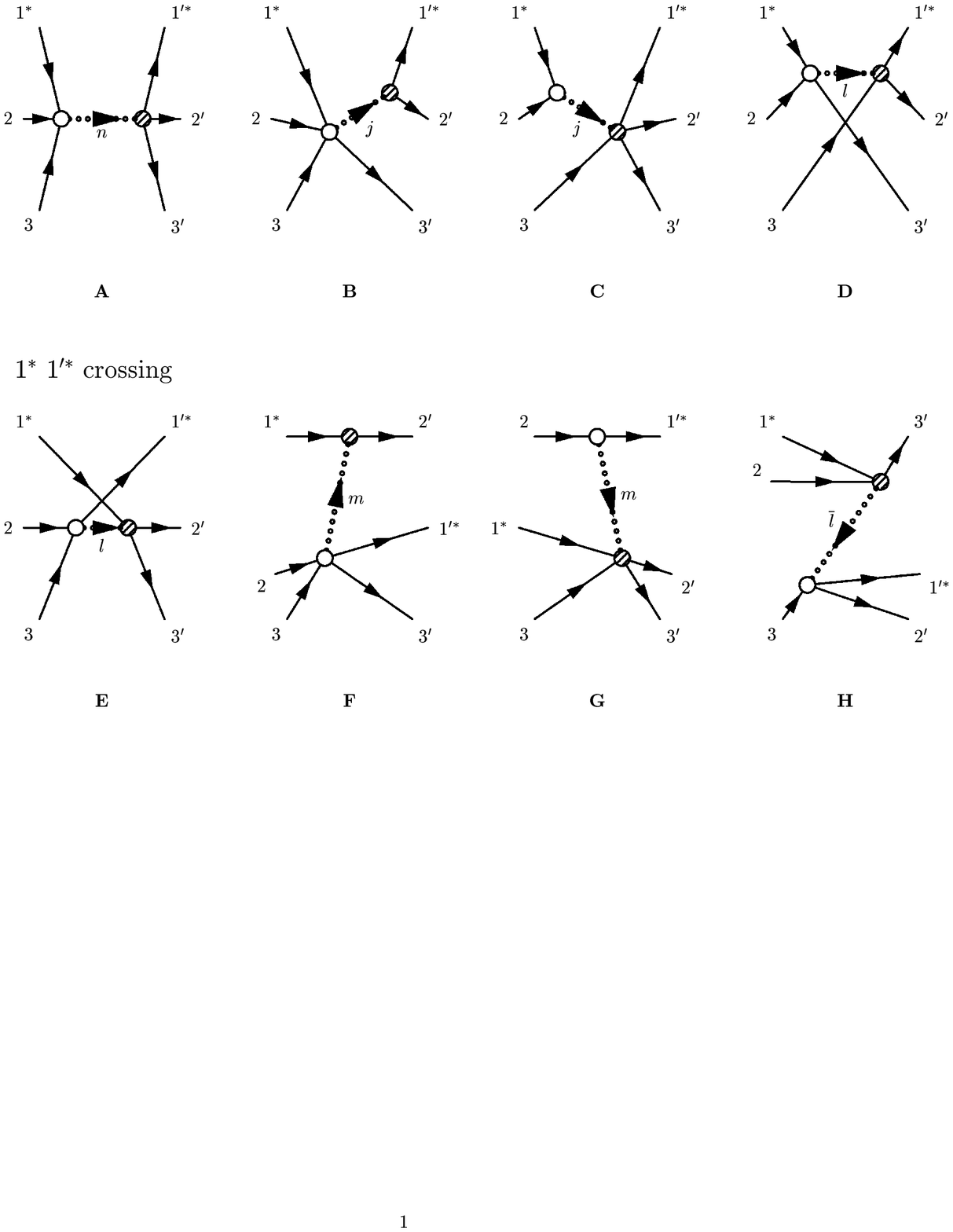} } 
\vspace{-5.0cm} 
\caption{{\protect\footnotesize {\it The  on mass shell particle
 exchange diagrams which are included
in the effective potential (3.11) for
the three-fermion reactions $1+2+3\to 1'+2'+3'$.
The empty circle stands for the primary
transition amplitude and the dashed circle corresponds to the following
transition amplitude. Fermions with the index $^*$ are extracted from
the asymptotic states 
in the expression (3.11) or in (3.8) $a\equiv 1'$ and  $b\equiv 1$. 
For any
amplitude in the left side of Eq. (3.11) or (3.8) only one particle 
 ($a$ or $b$) is 
considered off mass shell.  
All of diagrams have the three-dimensional
time-ordered form with the $^{\prime\prime}dressed^{\prime\prime}$
 vertices. Therefore 
in all of the diagrams the initial empty circle is depicted in left-hand
side and the following circle takes a place in the right-hand side. 
}}}
\label{fig:two}
\end{figure}
\vspace{5mm}

Equation (3.7) contains the auxiliary amplitude

$${\cal T}_{\alpha\beta}=-<in;{\widetilde \alpha}| J_{\bf p_a}(0)|\beta; in>
.\eqno(3.9)$$

In the transition matrix
$<{\widetilde \alpha}|J_{\bf p_a}(0)|\beta>$
with an arbitrary ${\widetilde \alpha}$ and $\beta$ states, all particle
except of $a$ are on mass shell. The four-momentum 
of particle $a$  is expressed through the four-momenta of
other on mass shell particles, i.e. $p_a=P_{\beta}-P_{\widetilde \alpha}$.
  Therefore afterwards we shall consider
particle $a$  as off mass shell particle 
in the corresponding matrix element.

For the one-particle asymptotic state ${\widetilde \alpha}\equiv 1'$ 
 we have 
$f_{1'+a,\beta}={\cal T}_{1'+a,\beta}$, because
$<out;1'|=<in;1'|$.  But for the three-particle asymptotic state
$<out;\alpha|$ the case is more complicate 
${\cal T}_{1'+2'+a,\beta}\ne{\cal T}_{1'+2'+a,\beta}$.
We can obtain the analogical to (3.7) relation
for ${\cal T}_{\alpha\beta}$ (3.9)
using the $S$-matrix reduction formulas  

$$
{\cal T}_{\alpha\beta}= w_{\alpha\beta}+
(2\pi)^3\sum_{\gamma} {\cal T}_{\alpha\gamma} {%
\frac{{\delta^{(3)}( {\bf p}_b+{\bf P_{{\widetilde \beta}}-P_{\gamma} })}}{{%
E_{\bf p_b}+{\ P_{{\widetilde \beta}}^o-P_{\gamma}^o+i\epsilon }} }} 
{{\cal T}^{\ast}}_{\beta \gamma} \eqno(3.10)
$$

\newpage

\vspace{5mm}


\begin{figure}[htb]
\centerline{\epsfysize=165mm\epsfbox{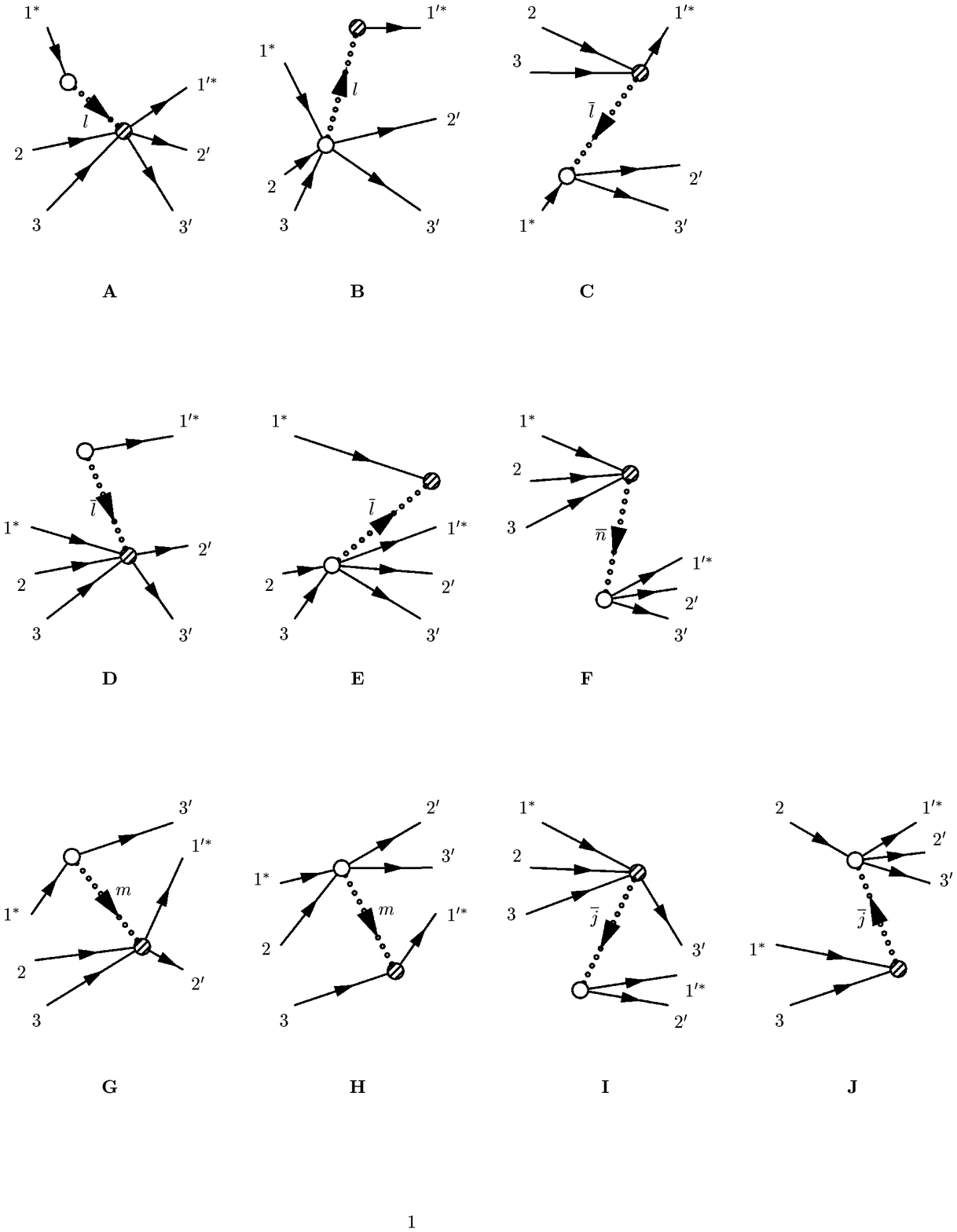}} 
\vspace{-1.0cm} 
\caption{{\protect\footnotesize {\it 
Diagrams obtained after the two particle $(2,3)$ and  $(2',3')$
transposition from the $s$-channel diagram in Fig.2A and from the 
$t$-channel diagram in Fig.2E.
}}}
\label{fig:three}
\end{figure}
\vspace{5mm}

where

$$
w_{\alpha\beta}= 
-<in;{\widetilde \alpha}|\Bigl\{J_{\bf p_a}(0),b_{\bf p_b}^+(0)\Bigr\} 
|{\widetilde \beta};in> $$
$$+(2\pi)^3\sum_{n=1''2''3''b'',...}<in;{\widetilde\alpha} |
J_{\bf p_a}(0)|n;in>
{ {\delta^{(3)}( {\bf p}_b+{\bf P}_{\widetilde \beta}-{\bf P}_{n} ) }\over
{E_{\bf p_b}+P_{\widetilde \beta}^o-P_{n}^o+i\epsilon } }
<in;n|{\overline J}_{\bf p_b}(0)|{\widetilde \beta};in>
$$
$$
-(2\pi)^3\sum_{l=f,fb,...} 
<in;{\widetilde \alpha}| {\overline J}_{\bf p_b}(0)|l;in>
 { {\delta^{(3)}( {-{\bf p}}_b+{\bf P_{\widetilde \alpha}-P_{l} }) } 
\over{-E_{\bf p_b}+ P_{{\widetilde \alpha}}^o-P_{l}^o } }
<in;l|J_{\bf p_a}(0)|{\widetilde \beta};in>,\eqno(3.11)
$$

Using Eq. (3.7) and Eq.(3.9) we can find the connections formula 
between $f_{\alpha\beta}$ and  ${\cal T}_{\alpha\beta}$. 
If we suppose, that
$${\cal T}_{\alpha\beta}=\Bigl(wW^{-1}f\Bigr)_{\alpha\beta}\eqno(3.12a)$$

and

$$f_{\alpha\beta}=\Bigl(Ww^{-1}{\cal T}\Bigr)_{\alpha\beta},\eqno(3.12b)$$
then after insertion of relation (3.12a) into  Eq.(3.7) we 
obtain Eq.(3.9) for 
${\cal T}_{\alpha\beta}$. And  vice versa,
after insertion of relation (3.12b) into Eq. (3.9) we get
Eq. (3.7). This is the justification of the
relations (3.12a,b) for the nonsingular effective three-body potentials
$W$ (3.8) and $w$ (3.11).

The consistent procedure of separation of the complete set 
of a connected  and disconnected parts
in the  three-dimensional equations  (3.10) or (3.7)
is well known as field-theoretical
cluster decomposition procedure\cite{alf,Ban}. 
In ref. \cite{M7} this procedure is
applied to the three-body equation for the  $\gamma\pi N$ systems. 
For the three-body reactions
$1+2+3\to 1'+2'+3'$ the cluster decomposition procedure is the same as
 separation of the following connected and disconnected matrix elements

$${\cal T}_{1'2'3',123}^{dc}=
-<in;{\bf p'_2,p'_3}|b_{\bf p_3}^+(in) J_{\bf p'_1}(0)|{\bf p_1,p_2}; in>$$
$$+<in;{\bf p'_2,p'_3}|b_{\bf p_2}^+(in) J_{\bf p'_1}(0)|{\bf p_1,p_3}; in>
-<in;{\bf p'_2,p'_3}|b_{\bf p_1}^+(in) J_{\bf p'_1}(0)|{\bf p_2,p_3}; in>$$
$$+<in;{\bf p'_2}| J_{\bf p'_1}(0)b_{\bf p'_3}(in)|{\bf p_1,p_2,p_3}; in>
-<in;{\bf p'_3}| J_{\bf p'_1}(0)b_{\bf p'_2}(in)|{\bf p_1,p_2,p_3}; in>
,\eqno(3.13)$$

$${\cal T}_{1'2'3',123}^{c}=-\sum_{permutations\ 1,2,3}
<in;{\bf p'_2,p'_3}|\biggl\{\Bigl\{ 
J_{\bf p'_1}(0),b_{\bf p_1}^+(in)\Bigr\},b_{\bf p_2}^+(in)
\biggr\}|{\bf p_3}; in>.\eqno(3.14)$$

The $s$ and $u$ channel terms of the effective potential (3.11)
are depicted in Fig.2A and in Fig.2E.
As off mass shell one-particle states  are taken
$b=1$ and $a=1'$. 
These off mass shell 
particle in the following figure are marked by $*$.
The diagrams in Fig.2A and in Fig.2E
have different chronological sequences of a absorption 
and a emission of the  particles $1$ and $1'$. 
In particular, the $s$-channel
diagram 2A corresponds to the chain of reactions, where firstly the initial
three body state $1^*+2+3$ transforms into intermediate on mass shell 
$n$-particle states which
afterwards produces the final  ${1^*}'+2'+3'$
 state. In the diagram 2E at first the final fermion 
${1^*}^{\prime}$ is generated with the intermediate states $l$
from the initial $2+3$ states and 
afterwards we obtain
final $2'+3'$ state from the intermediate $l+1^*$
states.

Using the cluster decomposition procedure for the $s$ and $u$ channel terms
in Eq. (3.11) or in Eq. (3.16) one can change 
the chronological sequence of absorption of the initial on mass shell 
fermions $2,3$ and 
emission of the final on mass shell particles 
$2',3'$. In diagrams 2B,2C and 2D
are presented all possible transpositions of the particles $3$ and $3'$
from the original $s$-channel diagram 2A which can be performed 
after transposition of particles $3$ and $3'$
using the disconnected structure (3.13) of the tree-body amplitudes.   
In particular, Fig.2B is obtained after transposition of 
fermion $3'$ i.e. after substitution of the disconnected part of amplitude 
$<in;{\bf p'_2}|J_{\bf p'_1}(0)b_{\bf p'_3}(in)|n; in>$.  Fig.2C is generated 
by transposition of fermion $3$
and Fig. 2D is result of the permutation of a both particles $3$ and $3'$.
Unlike to the diagram 2A, in the diagram 2B the
intermediate $l$ states arise 
together with the final $3'$ state and next
 are generating the
final two-fermion ${1^*}'+2'$ states. The same procedure of 
transposition of particles $3$ and $3'$ from the 
$u$-channel diagram in Fig.2E generates the diagrams 2F. 
2G and 2H. An other kind of permutations of the both particles
$2+3$ and $2'+3'$ from  $s$-channel diagram in Fig. 2A produces the 
diagrams 3A, 3B and 3C. In particular, 
diagrams 3G and 3I are obtained from diagrams 2B and 2F after transposition 
of the $2+3$ particle states. And  transposition of $2'+3'$ states
in diagrams 2C and 2G produces  3H and 3J diagrams.
The complete set of the diagrams which can be obtained after transpositions 
of the particles $(2,3);(2'3')$ consists from the different disposition
of these particles at the first vertex function
and at the following vertex function.
The first vertex function in Fig. 2 and in Fig.3
is denoted with the empty circle in Fig. 2 and 
the dashed circle stands for the next vertex function.
One has the following combinations of the dispositions
of particles  $(2,3);(2'3')$ at the vertex functions:
$1^*+zero\ particles\ {\Longrightarrow}{1'}^*+four\ particles$, 
$1^*+one\ particle\ {\Longrightarrow}{1'}^*+three\ particles$, 
$1^*+two\ particle\ {\Longrightarrow}{1'}^*+two\ particles$, 
 $1^*+three\ particles\ {\Longrightarrow}{1'}^*+one\ particle$ and 
 $1^*+four\ particles\ {\Longrightarrow}{1'}^*+zero\ particles$. 
For instance, we have four diagrams  
$1^*+2\ {\Longrightarrow}{1'}^*+3,2'3'$ (Fig. 2C),
$1^*+3\ {\Longrightarrow}{1'}^*+2,2'3'$,
$1^*+2'\ {\Longrightarrow}{1'}^*+23,3'$ and
$1^*+3'\ {\Longrightarrow}{1'}^*+23,2'$ (Fig. 3G) for the disposition
 $1^*+one\ particle\ {\Longrightarrow}{1'}^*+three\ particles$.
The particle distribution
$1^*+three\ particles\ {\Longrightarrow}{1'}^*+one\ particle$
have also the four diagrams
${1}^*+3,2'3'\ {\Longrightarrow}{1'}^*+2$,
${1}^*+2,2'3'\ {\Longrightarrow}{1'}^*+3$ (Fig. 3H),
${1}^*+23,3'\ {\Longrightarrow}{1'}^*+2'$ (fig. 2B),
${1}^*+23,2'\ {\Longrightarrow}{1'}^*+3'$.
The particle distribution
$1^*two\ particle\ {\Longrightarrow}{1'}^*+two\ particles$
can be observed  in  the six diagrams
 $1^*+23{\Longrightarrow}{1'}^*+2'3'$ (Fig.2A),
 $1^*+2,3'{\Longrightarrow}{1'}^*+3,2'$ (Fig. 2D),
 $1^*+2,2'{\Longrightarrow}{1'}^*+3,3'$,
 $1^*+3,2'{\Longrightarrow}{1'}^*+2,3'$,
 $1^*+3,3'\ {\Longrightarrow}{1'}^*+2,2'$ and
 $1^*+2'3'{\Longrightarrow}{1'}^*+23$ (Fig. 3C).
And one diagram  
 $1^*\ {\Longrightarrow}{1'}^*+23,2'3'$ (Fig. 3A) and one diagram
 $1^*+23,2'3'\ {\Longrightarrow}{1'}^*$  (Fig.3B) relates  to the
distributions
$1^*+zero\ particles\ {\Longrightarrow}{1'}^*+four\ particles$
and 
$1^*+four\ particles\ {\Longrightarrow}{1'}^*+zero\ particles$
correspondingly.
Thus the $s$-channel term in Eq.(3.16) generates the 
 $2\times 4+6+2=16$ connected terms
 after cluster decomposition.
The other 16 connected terms  produces the $u$-channel term in Eq.(3.16). 
Thus we get 32 independent skeleton 
 diagrams after the cluster decomposition procedure performed in 
the second and in the third terms of Eq.(3.11) or Eq.(3.16).
Diagrams 3C, 3D, 3E, 3F, 3I and 3J contain the antiparticle intermediate 
states,
because the time-ordered field-theoretical formulation 
includes the complete set 
of the intermediate particle propagators with a 
different time sequences.
This means that for  any diagrams with $n,l,...$-particle intermediate 
states appear the corresponding diagrams with the antiparticle 
${\overline n},{\overline l},...$ intermediate states.

The $S$-matrix reduction formulas for the 
$1d\Longrightarrow 1'+d'$ process gives us the following equation

$$
{\cal T}_{1'd',1d}= 
-<out;{\bf P'_d}| J_{\bf p'_1}(0)|{\bf p_1,P_d};in>
=w_{1'd',1d}+
(2\pi)^3\sum_{n=3f,fd,d3} {\cal T}_{1'd',\gamma}
{{\delta^{(3)}( {\bf p}_1+{\bf P_{d}-P_{\gamma} })}\over
{E_{\bf p_1}+P_{d}^o-P_{\gamma}^o+i\epsilon } }  
{\cal T}^{\ast}_{1d,\gamma} \eqno(3.15)
$$

where

$$
w_{1'd',1d}= 
-<out;{\bf P'_d}|\Bigl\{J_{\bf p'_1}(0),b_{\bf p_1}^+(0)\Bigr\} 
|{\bf P_d};in> $$
$$+(2\pi)^3\sum_{n=1''2''3''b'',...}<out;{\bf P'_d} |
J_{\bf p'_1}(0)|n;in>
{ {\delta^{(3)}( {\bf p}_1+{\bf P}_{d}-{\bf P}_{n} ) }\over
{E_{\bf p_1}+P_{d}^o-P_{n}^o+i\epsilon } }
<in;n|{\overline J}_{\bf p_b}(0)|{\bf P_d};in>
$$
$$
-(2\pi)^3\sum_{l=f,fb,...} 
<out;{\bf P'_d}| {\overline J}_{\bf p_1}(0)|l;in>
 { {\delta^{(3)}( {-{\bf p}}_b+{\bf P_{\widetilde \alpha}-P_{l} }) } 
\over{-E_{\bf p_1}+ P_{d'}^o-P_{l}^o } }
<in;l|J_{\bf p'_1}(0)|{\bf P_d};in>.\eqno(3.16)
$$

\vspace{5mm}



\begin{figure}[htb]
\centerline{\epsfysize=145mm\epsfbox{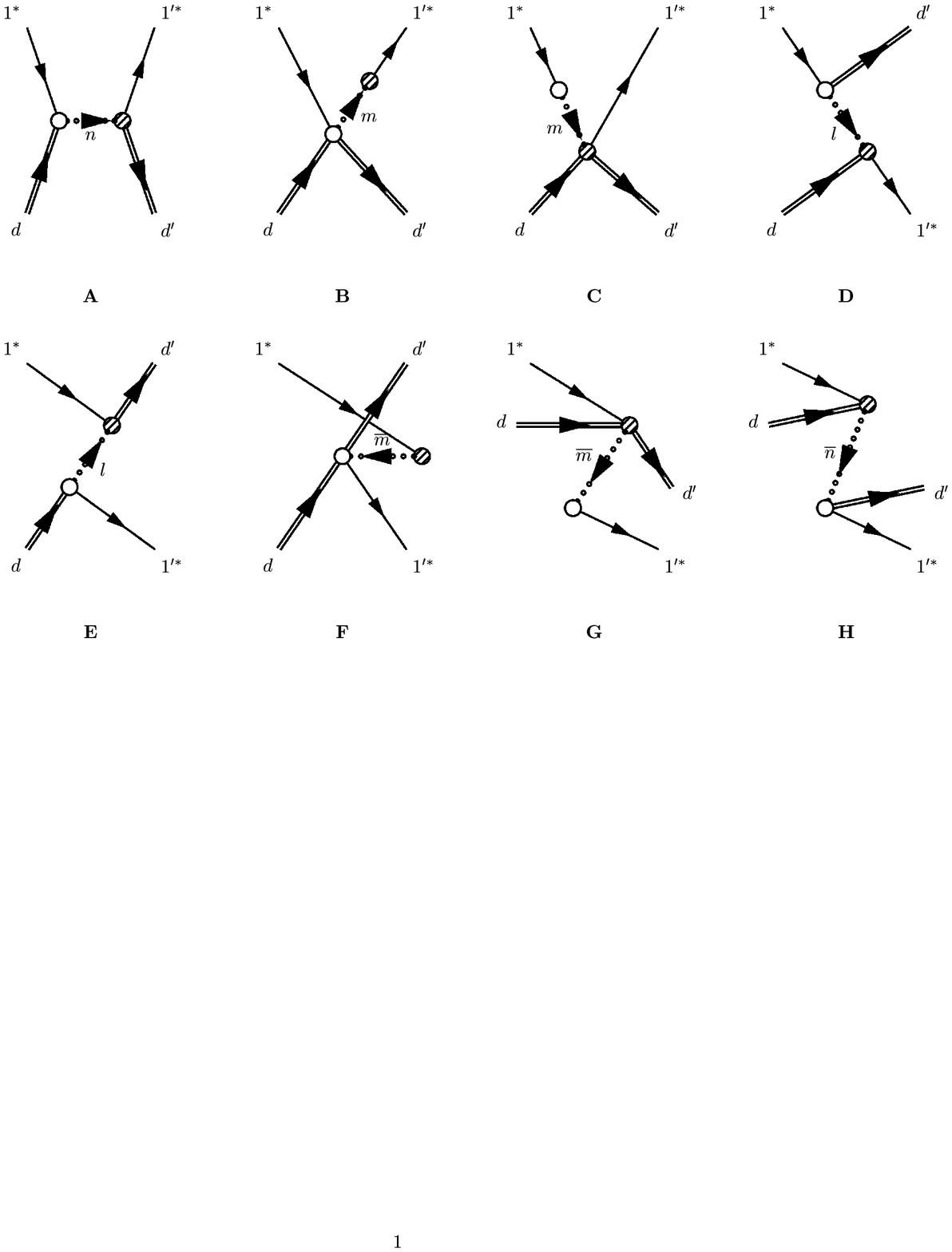}} 
\vspace{-7.0cm}
\caption{{\protect\footnotesize {\it The graphical representation of the 
on mass shell particle exchange
potential (3.16) for the $1+d\Longrightarrow 1'+d'$ amplitude
 after cluster decomposition.
The double line denotes a two-fermion bound state, $n$ stands for the 
three-fermion$+boson$ states, $m=f+b,...$ and $l$ corresponds to the 
intermediate fermion $f,f+b,...$ states.
}}}
\label{fig:four}
\end{figure}
\vspace{3mm}

Equations (3.15) and (3.16) for the amplitude and for the effective
potential of the 
$1d$ scattering reaction
 have the same form as the two-body equations.
After cluster decomposition of $w_{1'd',1d}$ we obtain only 6 terms 
(see Fig.4) with a transposition $d\Longleftrightarrow d'$ and 
with crossing transformation $1'\longleftrightarrow 1$. 
Eq.(3.15) contains the 
$1d\to 1'2'3'$ transition matrix which is connected with  the
$123\to 1'2'3'$ transition amplitude according to  the Eq.(3.10),
 where $\alpha,\beta=3f$, but $\gamma=3f,fd$.
Thus the  $123\to 1'2'3'$, $1d\to 1'2'3'$ and $1d\to1'd'$ transition
amplitudes are the solutions of the coupled equations
(3.10) and (3.15). Using the linearization procedure of such type equations
 \cite{M1,M7}, one can obtain the equivalent set of 
Lippmann-Schwinger-type equations

$$
{\cal T}_{\alpha,fd}(E_{fd})=U_{\alpha,fd}(E_{fd})+
\sum_{\gamma} U_{\alpha\gamma}(E_{fd}) { 1\over
{ E_{fd} -E_{\gamma}+i\epsilon } } {\cal T}_{
\gamma,fd}(E_{fd}),\eqno(3.17)
$$

where $E_{fd}=E_d+E_f$ is the energy of the asymptotic fermion and 
two-fermion bound state $d$, $\alpha,\gamma=3f,fd$ and $U_{\alpha\gamma}(E)$
is unambiguously determined from the connected part of potentials 
$w_{\alpha\gamma}$ according to 
the relation (2.9a). Note that the solution of the three-body equations
(i.e the $123\to 1''2''3'$ and $123\to 3'd''$ transition amplitudes)
 participate in the 
$w_{\alpha\gamma}^c$ potential in the diagrams 3B and 3C.
One can rid the three-fermion potential
of such type nonlinearities after introduction of a new 
amplitudes
$f_{\alpha,\beta}=F_{\alpha,\beta}+A_{\alpha,\beta}$ 
in Eq.(3.10) and in Eq.(3.15), where  the choice of 
$A_{\alpha,\beta}$ is conditioned by 
 cancellation of the terms in Fig. 2B and in Fig. 2C which have
the form $fg_oA^+$ and $Ag_of^+$.
Afterwards we get the linear Lippmann-Schwinger-type equation 
for $ F_{\alpha,\beta}$ amplitudes with the disconnected terms.
Therefore this linearization procedure generates necessity to use 
the Faddeev-type equations for the three-fermion scattering problems.
Certainly, in the intermediate energy region, i.e
 up to $2GeV$ of the energy of the incoming proton for the $Nd-3N$ systems,
one can neglect the diagrams 3I and 3J with the $2{\overline N}$ states
and the diagram 3F with the $3{\overline N}$ intermediate state.

Equations (3.10) 
and (3.15) represent the spectral decomposition
formulas (or off shell unitarity conditions) for the three-body amplitudes
in the standard quantum field theory. Such three-dimensional time-ordered
relations ware  considered in the textbooks in the quantum field theory 
\cite{BD,IZ,alf} and in the nonrelativistic collision theory \cite{New,Gold}
for the two-body reactions. Therefore, one can treat Eq. (3.10) and Eq.
(3.15) as the three-body generalization of the field-theoretical
spectral decomposition formulas (or off shell unitarity conditions) for the
two-body amplitudes.
The field-theoretical formulation allows us to obtain the analytical
structure of the three-body amplitudes and
the problem of determination of the three-body forces in the nonrelativistic
Faddeev equations does not arise in the considered formulation.


\medskip

\begin{center}
{\bf 4.  Equal-time anticommutators as a off mass shell particle
exchange potential.}
\end{center}

\medskip

The important part of the effective potential $w_{\alpha\beta}$ is 
the equal-time commutator in the effective potentials (3.11) and (3.16).
The equal-time anticommutators 
in the effective potential of the 
$1d\to 1'd'$, $1d\to 1'2'3'$ and $123\to 1'2'3'$ reactions are
$$
Y_{1d,1'd'}=
-<{\bf P'_d}|\Bigl\{J_{\bf p'_1}(0),b_{\bf p_1}^+(0)\Bigr\} 
|{\bf P_d};in>.\eqno(4.1a)
$$

$$
Y_{1d,1'2'3'}=
-<{\bf p'_2,p'_3}|\Bigl\{J_{\bf p'_1}(0),b_{\bf p_1}^+(0)\Bigr\} 
|{\bf P_d};in>.\eqno(4.1b)
$$

$$
Y_{123,1'2'3'}=
-<{\bf p'_2,p'_3}|\Bigl\{J_{\bf p'_1}(0),b_{\bf p_1}^+(0)\Bigr\} 
|{\bf p_2,p_3};in>.\eqno(4.1c)
$$


\begin{figure}[htb]
\centerline{\epsfysize=145mm\epsfbox{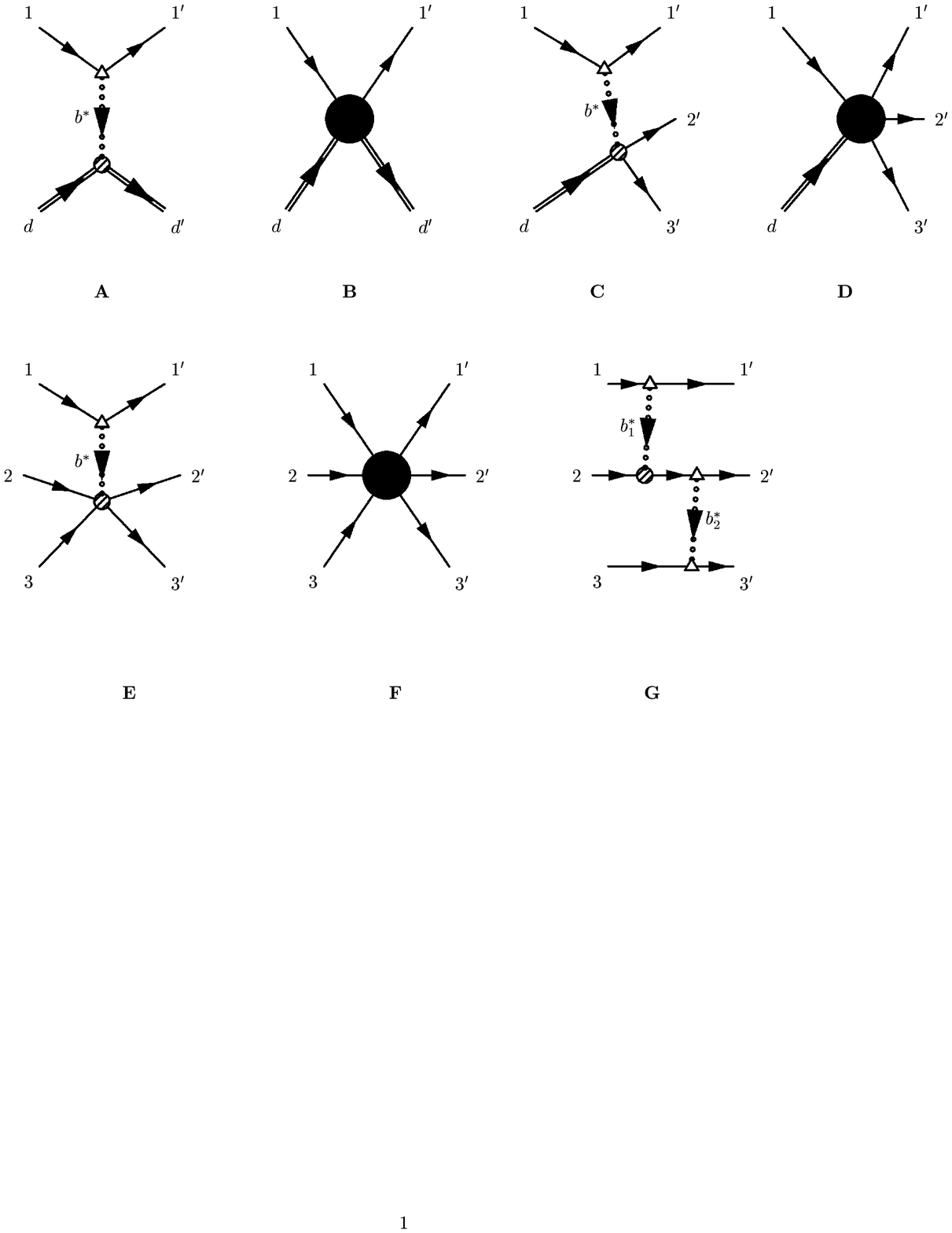}} 
\vspace{-6.0cm}
\par
\caption{{\protect\footnotesize {\it The graphical representation of the
equal-time anticommutators (4.1a,b,c). These terms
are depicted separately for the binary reactions ${\bf A,B}$, for the
process $2\Longrightarrow 3^{\prime}$ ${\bf C,D}$ and for the three-body
process $3\Longleftrightarrow 3^{\prime}$ ${\bf E,F,G}$. 
Diagrams ${\bf A,C,E}$ 
correspond to the one off-mass shell particle exchange interactions
which are appearing from in the $\phi^3$-theory i.e. for the QED or 
for the Yukawa-type interactions. The triangle denotes the vertex functions
in the tree approximation.
Diagrams ${\bf B,D,F}$ 
describe the contact (overlapping) interaction which does not
contain the intermediate hadron propagation between hadron states.
Diagram 5G corresponds to the simplest one off mass shell boson fermion
and two off mass shell boson exchange 
interaction which is obtained from the equal-time 
commutators in Eq.(4.1c) (diagram 5F) in the framework of the $\phi^3$
theory. }}}
\label{fig:five}
\end{figure}
\vspace{3mm}

The explicit form of expression (4.1abc) can be determined using 
the $a\ priori$ given Lagrangian 
and equal-time anticommutations relation between the Heisenberg
field operators. In the case of renormalizable Lagrangian models or for
nonrenormalizable simple phenomenological Lagrangians the equal-time
anticommutators are easy to calculate \cite{M1,M7}. 
In that case  expressions (4.1abc) which often are called 
as the seagull terms,
consists of the off shell ${\bf internal}$ one particle exchange potentials
(see diagrams 5A, 5C and 5E) and of the contact (overlapping) terms
(Fig.5B, Fig.5D and Fig.5F) which does not contain any particle
propagator in the intermediate states between 
asymptotic $|{\overline \beta}>$ and $<{\overline \alpha}|$ states.
The equal-time commutators are the only  part of 
effective potentials (3.11) or (3.16) which
contains ${\bf explicitly}$ the ${\bf internal}$ off mass shell 
particle exchange
diagrams, since other terms in the effective potential (3.11)  or (3.16)
consists of the 
on mass shell particle exchange terms,
where off mass shell are ${\bf external}$ fermions. In order to
clarify the structure of the equal-time terms, we will consider Lagrangian
of the simplest $\phi^3 $ model for the electromagnetic fields and for the 
psevdoscalar $\pi N$ interactions

$${\cal L}_{em}= -e{\overline \psi}\gamma^{\mu}\psi A_{\mu};\ \ \
 {\cal L}_{ps}=-ig_{\pi}{\overline \Psi}\gamma_5 \tau \Phi_{\pi}\Psi.
\eqno(4.2)$$

The current operator and the equation of motion for this Lagrangians
 are

$$
\partial_{\nu}\partial^{\nu}A_{\mu}=J_{\mu}= -e{\overline \psi}\gamma^{\mu}
\psi;\ \ \ 
(\partial_{\nu}\partial^{\nu}+m_{\pi}^2)\Phi_{\pi}^i=j_{\pi}^i= -ig_{\pi}
{\overline \Psi}\gamma_5 \tau^i\Psi\eqno(4.3)
$$

Using  the equal-time anticommutations relation between the Heisenberg field 
operators for the expressions(4.1abc)
 we get 

$$
Y_{\alpha\beta}= 
{-e\over {\sqrt{Z_1Z_{1'} } }}
 {\overline u}({\bf p'_1})\gamma^{\mu}u({\bf p_1})
<in;{\widetilde \alpha}|A_{\mu}(0)|{\widetilde \beta};in>=
{-e\over {\sqrt{Z_1Z_{1'} } }}
 {{  {\overline u}({\bf p'_1})\gamma^{\mu}u({\bf p_1}) }
\over{(P_{\widetilde \alpha}-P_{\widetilde \beta})^2 } } 
<in;{\widetilde \alpha}|J_{\mu}(0)
|{\widetilde \beta};in>,\eqno(4.4a)$$
or for the $\pi NN$ system

$$Y_{\alpha\beta}= 
{ {-ig_{\pi}}\over{\sqrt{Z_1Z_{1'}} } }
{\overline u}({\bf p'_1})\gamma^{5}\tau^iu({\bf p_1})
<in;{\widetilde \alpha}|\phi_{\pi}^i(0)
|{\widetilde \beta};in>=
  { {-ig_{\pi}}\over {\sqrt{Z_1Z_{1'} } } }
{ {  {\overline u}({\bf p'_1})\gamma^5\tau^i u({\bf p_1}) }
  \over{(P_{\widetilde \alpha}-P_{\widetilde \beta})^2-m_{\pi}^2 } } 
<in;{\widetilde \alpha}|j_{\pi}^i(0)
|{\widetilde \beta};in>,\eqno(4.4b)$$

where ${\widetilde \alpha},{\widetilde \beta}=
(d,d'),(d,2'3')\ and\ (23,2'3')$ for (4.1a), (4.1b)
 and (4.1c) correspondingly. Expressions (4.4a) or (4.4b) 
relates to the one off mass shell boson 
exchange diagrams 5A, 5C and 5E for the Lagrangians (4.2).
Using more complete Lagrangian models, 
one can obtain also heavy $\rho,\omega$
meson exchange diagrams \cite{M1,M7}. Moreover, in the ref.\cite{M1} the
One Boson Exchange (OBE) Bonn model of $NN$ potential was exactly
reproduced from the equal-time anticommutators. 
There was also numerically  estimated the contributions from the contact 
(overlapping) terms for $NN$ phase shifts. 
These contact (overlapping) terms arise
from the $\phi^4$ (four-point) part of Lagrangians or from the
nonrenormalizable Lagrangians and they play an 
important role for the $NN$ scattering.

The other source for the overlapping (contact) terms
in the quantum field theory is 
the quark-gluon degrees of freedom.
One can
construct the hadron creation and annihilation operators as well as the
Heisenberg field operators of hadrons from the quark-gluon fields in the
framework of the Haag-Nishijima-Zimmermann \cite{HNZ} treatment
of the composed particles. In
this case  the contact terms (see diagrams 5B, 5D and 5F) contains the 
contributions from the quark-gluon exchange \cite{M1,MBFE}.
But the  
equations (3.10), (3.11), (3.15) and (3.16) remain be the same\
also for the formulation with the quark-gluon degrees of freedom.

The contact (overlapping) terms depicted in 
diagrams 5C, 5D and 5E, 5F
can be treated as pure
three-body forces. For these terms 
 it is necessary to use an additional derivation of two-body and
the three-body
equations like the spectral decomposition formulas (3.10). 
These extra auxiliary two-body and the 
three-body amplitudes are necessary for solution of the 
considered three-body equations.
As example we shall consider the amplitude  
$<{\bf p'_2,p'_3}|j_b(0)|{\bf p_2,p_3}>$ for the reaction $23\to b'2'3'$. 
 This amplitude participate in the diagram 5E and the corresponding
on mass shell particle exchange diagrams are depicted in Fig.6.
Diagrams 6B, 6C, 6D are obtained from the $s$-channel diagram 6A after
transpositions $2\Longleftrightarrow 1'$. The next four diagrams are
 produced by crossing permutation of the off mass shell particles $1^*$ and
${b'}^*$. The last four diagrams are obtained from diagrams 6A, 6C, 6E and 6C 
after transposition of the $1'2'$ states to the first vertex.

The complete set of the diagrams which can be obtained 
after transpositions of the particles $2;1'2'$ 
in the $s$ channel term of amplitude of the
 $1^*+2\Longrightarrow{b'^*}1'2'$ reactions
consists from the following dispositions  of on mass shell
particles $2;(1'2')$ at the first and the next vertices:
$1^*+zero\ particles\ {\Longrightarrow}{b'}^*+three\ particles$, 
$1^*+one\ particle\ {\Longrightarrow}{b'}^*+two\ particles$, 
$1^*+two\ particle\ {\Longrightarrow}{b'}^*+two\ particles$ and  
$1^*+three\ particles\ {\Longrightarrow}{b'}^*+zero\ particles$. 
For instance, we have three diagrams 
$1^*+2\ {\Longrightarrow}{b'}^*+1'2'$ (Fig.6A),
$1^*+1'\ {\Longrightarrow}{b'}^*+2,2'$(Fig. 6D) and
$1^*+2'\ {\Longrightarrow}{b'}^*+2,2'$ 
for the distribution
 $1^*+one\ particle\ {\Longrightarrow}{b'}^*+two\ particles$.
The particle dispositions 
$1^*+two\ particles\ {\Longrightarrow}{b'}^*+one\ particle$
can be realized also with the three diagrams
${1}^*+2,1'\ {\Longrightarrow}{b'}^*+2'$,
${1}^*+2,2'\ {\Longrightarrow}{b'}^*+1'$ (Fig. 6B) and
${1}^*+1'2'\ {\Longrightarrow}{b'}^*+2$ (Fig. 6J).
And the diagrams with the particle distribution 
${1}^*+2,1'2'\ {\Longrightarrow}{b'}^*$ (Fig. 6I) and 
${1}^*\ {\Longrightarrow}{b'}^*+2,1'2'$ (Fig. 6C) form the 
distributions
$1^*+three\ particles\ {\Longrightarrow}{b'}^*+zero\ particles$
 and
$1^*+zero\ particles\ {\Longrightarrow}{b'}^*+three\ particles$. 
Therefore the complete set of the connected  $s$-channel terms
 are $2\times 3 +2=8$.  Together with the $u$ channel terms we 
have 32 diagrams for the $b$-boson creation reaction on the two fermion 
system.

The simplest contact terms (4.1c) for the three-point Lagrangians (4.2)
have the form

$$
Y_{1'2'3',123}= 
-e^3 {{  {\overline u}({\bf p'_1})\gamma^{\mu}u({\bf p_1}) }
\over{(P_{2'+3'}-P_{2+3})^2 } } 
<{\bf p'_2}|J_{\nu}(0)|{\bf p_2}>
{{  {\overline u}({\bf p'_3})\gamma^{\nu}(\gamma^{\sigma}Q_{\sigma}+m_{el})
\gamma_{\mu}u({\bf p_3}) }
\over{Z_1Z_3(p'_3-p_3)^2(Q^2-m_{el}^2)}}
,\eqno(4.5a)$$
and for the $\pi NN$ system

$$
Y_{1'2'3',123}= 
-i^3g_{\pi}^3 {{  {\overline u}({\bf p'_1})\gamma^{5}\tau^iu({\bf p_1}) }
\over{(P_{2'+3'}-P_{2+3})^2-m_{\pi}^2 } } 
<{\bf p'_2}|j_{\pi}^k(0)|{\bf p_2}>
{{  {\overline u}({\bf p'_3})\gamma^{5}\tau^k
(\gamma^{\sigma}Q_{\sigma}+m_{N})
\gamma_{5}\tau^iu({\bf p_3}) }
\over{Z_1Z_3\Bigl((p'_3-p_3)^2-m_{\pi}^2\Bigr)  (Q^2-M_N^2)}}
,\eqno(4.5b)$$

where $Q=p_1+p_2-p'_1$ and this simplest two off mass shell 
boson exchange term
is depicted in Fig. 5G.

Starting from any Lagrangian 
we  always obtain  one off mass shell boson exchange potentials 
(Fig. 5A, Fig.5C and Fig. 5E). The contact (overlapping) terms
we use $\phi^4$ terms in Lagrangian \cite{M7}, or more complicated 
models of phenomenological Lagrangians \cite{M1},
models of a nonrenormalizable Lagrangian, 
quark-gluon degrees of freedom \cite{M7} etc. 
These terms contains the other kind of a three-body amplitudes too and one 
must include these extra auxiliary two-body and 
 three-body amplitudes in the set of coupled equations
(3.10) and (3.15). Thus the number of the solved three-body equations
and the form of the auxiliary amplitudes  
is depending on the form of ``input'' Lagrangian. This means,
 that the unified description of the coupled 
three-fermion reactions 
can help us to determine the form of a
input Lagrangians which are
sufficient and necessary for a description of the experimental observables.

Besides of the Lagrangians, as ``input'' by construction of the 
effective three-body potentials are the amplitudes of the $2\to 2'$
and $2\to 3'$ reactions and the three-point vertex functions.
In these vertex functions two particle are on mass shell 
and they are the function of a one variable. 
Therefore, one can determine these vertex functions from
the experimental data using
the quark counting rules, dispersion relations,
the Regge trajectories theory, or
inverse scattering method\cite{Ml}.



\begin{figure}[htb]
\centerline{\epsfysize=145mm\epsfbox{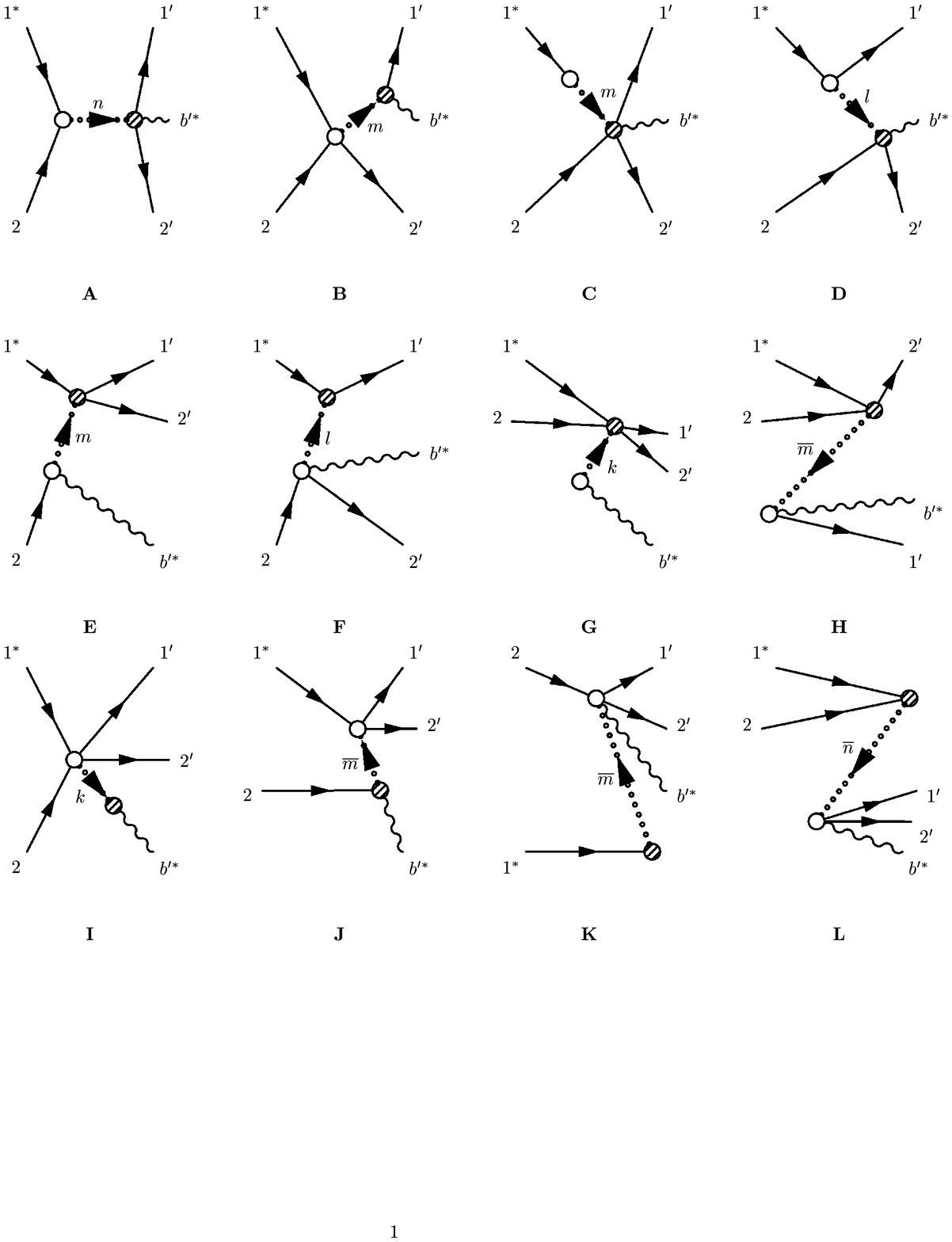}} 
\vspace{-2.0cm}
\caption{{\protect\footnotesize {\it The graphical representation of the 
on mass shell particle exchange
potential for the $1+2\Longrightarrow 1'+2'+b'$ amplitude
with off mass shell boson $b'$. 
This amplitude arise in the seagull term (4.1c)
in the $\phi^3$ theory after cluster decomposition.
The curled line denotes the off mass shell boson ${b'}^*$ 
 which corresponds to the photon 
$\gamma$ for leptons or 
$\pi$-meson for 
barions.
$n=2f''+b'',2f''+2b'',d''+b'',...$ 
stands for the intermediate on mass shell particle
states in $s$ channel, 
$m=f''+b'',...$ and $l=b'',2b'',..$.
}}}
\label{fig:six}
\end{figure}
\vspace{3mm}


\vspace{5mm}

\begin{center}
{\bf 5. Summary}
\end{center}

\vspace{5mm}

In my talk I have considered the three-dimensional covariant scattering
equations for  the amplitudes of the 
three-fermion scattering reactions. 
The basis of these three-body
relativistic equations is the standard field-theoretical $S$-matrix
reduction formulas. After decomposition over the complete set of the
asymptotic $^{\prime\prime}in^{\prime\prime}$ states the quadratically
nonlinear three-dimensional equations (3.10) or (3.15) were derived.
These equations have the same form as the off-shell unitarity conditions 
(2.1) in the nonrelativistic collision theory.
Afterwards 
the suggested nonlinear equations 
can be replaced by the equivalent Lippmann-Schwinger type equations
(2.2) or (3.17) for the connected part of the three-body amplitudes.
If one want to rid the potential of these three-body equations from
the nonlinear terms (see diagrams in Fig. 2B and in Fig. 2C with j=f''f''),
the after simple transformations 
one obtains the Lippmann-Schwinger-type equations with the usual
disconnected terms (see the last three diagrams in Fig.1C)
of a three-body potential. In this case instead of Eq. (3.17) 
we get the Faddeev-type equations.
Equations (3.17) satisfy all first principles of the quantum field theory. 
Moreover, as ``INPUT'' for construction of
the potential of these equations are required the one variable 
vertex functions which can be determined from the dispersion relations,
or from the inverse scattering method i.e. they can be obtained
from the two-body scattering observables.

The effective potential of the suggested equations consists
(i) from the on mass shell particle exchange diagrams in Fig.2, Fig.3
and Fig.4 and (ii) from the equal-time commutators  which
contains the one off-mass shell boson exchange diagrams (Fig. 5A,
Fig.5C and Fig.5D) and overlapping (contact) terms (Fig.5B, Fig. 5D
and Fig.5E).
The form and the number of this equal-time potentials
depend on the input Lagrangian model. 
For the three leptons interactions the overlapping (contact) terms does not 
appear and the diagram in Fig. 5E  are reduced to the simple off mass shell
 two-photon exchange diagram in Fig. 5F.  
In the case of  the two-body reactions
the   equal-time
commutators generates effective potential
which can be constructed from 
the phenomenological 
one variable vertex functions if one use the simple
the phenomenological
Lagrangians. These  one variable vertex functions may
be also determined from the 
 the experimental observables. 

In order to construct  the 
three-fermion potential from the one-variable phenomenological vertices 
one need  also to construct the two-fermion scattering amplitudes,
the two-fermion$\Longrightarrow$two-fermion$+$boson transition amplitudes
(see Fig.5E and Fig.6) and also the complicated overlapping (contact) term
 (Fig.5D and Fig.5F). 
The main attractive feature of the considered field-theoretical
scheme of the three-body equation is that it allows us to estimate
the importance of the overlapping (contact) terms.
Therefore the unified description of a two-body and a three-body
reactions in the considered formulation
allows us to determine the form of the simplest Lagrangians
which are necessary and enough for the unified 
description of the two-body and the three-body experimental data.
In addition these calculations
allows us to improve the accuracy of  the  calculations in the 
tree and in the Born approximations.

The considered field-theoretical formulation is not less general 
as the four dimensional Bethe-Salpeter equations. The final form of 
the equations (3.11) or (3.15) are not depending on the 
choice of the Lagrangian
and these equations are valid for any QCD motivated models with the 
quark-gluon degrees of freedom. But the suggested equations are much 
simpler as the 
analogical Bethe-Salpeter equations
and they can be numerically solved with the present computers. The only 
principal approximation,
that is necessary to do in this approach is the truncation of the 
intermediate multi-particle states. But here, unlike to the 
Bethe-Salpeter equations, one use to cut down only the
 ${\underline on\ mass\ shell\ intermediate\ states}$. In any case
 for the self-consistent calculation of the two=body and the three
body reactions in the low and intermediate energy region,
it is advisable to work out the scheme of a suppression  
mechanism of the transition of
{\cal one off mass shell fermion into on mass shell fermion$+$on 
mass shell boson} which arise together with the   
transition amplitude into the three-fermion$+$boson  states
 (see Fig.3A, Fig.3B,Fig.3D and Fig.3E).



\begin{thebibliography}{99}

\bibitem{Kvin}  A. N. Kvinikhidze and B. Blankleider, Nucl. Phys. {\bf A574}
(1994) 488.

\bibitem{Afnan}  D. R. Phillips and I.R. Afnan, Ann. of. Phys. {\bf 240}
(1995) 266 and {\bf 247} (1996) 19.


\bibitem{New}  R. G. Newton, Scattering Theory of Waves and Particles. (New
York-Heidelberg-Berlin, Springer-Verlag) 1982.

\bibitem{Gold}  M. L. Goldberger and M. Watson, Collision Theory. (New
York-London-Sydney, John Wiley and Sns.) 1965.

\bibitem{BD}  J. D. Bjorken and S.D.Drell, Relativistic Quantum Fields. (New
York, McGraw-Hill) 1965.


\bibitem{IZ}  C. Itzykson and C. Zuber. Quantum Field Theory. (New York,
McGraw-Hill) 1980.

\bibitem{GrossB}  F. Gross. Relativistic Quantum Mechanics and Quantum Field
Theory. (New York-London-Sydney, John Wiley and Sns. ) 1993.

\bibitem{M7}  A. I. Machavariani and Amand Faessler, Preprint 
arXiv:nucl-th/0306040 2003; to be publ. Ann. Phys.


\bibitem{alf}  V. De Alfaro, S. Fubini, G. Furlan and C. Rosseti, Currents
in Hadron Physics (North-Holland, Amsterdam) 1973.

\bibitem{Ban}  M. K. Banerjee and J. B. Cammarata, Phys. Rev. {\bf C17}
(1978) 1125.

\bibitem{M1}  A. I. Machavariani, Sov.J.Part. Nucl.
{\bf 24(3)} (1993) 731.


\bibitem{MBFE}  A. I. Machavariani, A. J. Buchmann, Amand Faessler, and G.
A. Emelyanenko, Ann. of. Phys. {\bf 253} (1997) 149.


\bibitem{Ml}  A. I. Machavariani, Phys. Lett. {\bf B540} (2002) 81.

\bibitem{HNZ}  R. Haag, Phys. Rev. {\bf 112} (1958) 669;
 K. Nishijima, Phys. Rev. {\bf 111} (1958) 995;
W. Zimmermann, Nuovo Cim. {\bf 10} (1958) 598;
K. Huang and H. A. Weldon, Phys. Rev. {\bf D11} (1975) 257.

\end{thebibliography}
\end{document}